\documentclass[prl,twocolumn,amsmath,amssymb,floatfix,groupedaddress]{revtex4-1}

\usepackage{graphicx}
\usepackage{amssymb}
\usepackage{amsmath}
\usepackage{rotating}
\usepackage{hyperref}

\usepackage[usenames,dvipsnames]{xcolor}

\DeclareMathAlphabet\mathbfcal{OMS}{cmsy}{b}{n}
\definecolor{darkgreen}{cmyk}{0.85,0.2,1.00,0.2} 
\definecolor{purple}{cmyk}{0.5,1.0,0,0}

\newcommand{\mnu}{M$\nu$}
\newcommand{\anu}{A$\nu$}
\newcommand{\snu}{S$\nu$}
\newcommand{\som}{S_8}
\newcommand{\Om}{\Omega_m}
\newcommand{\Oc}{\Omega_c}
\newcommand{\Ob}{\Omega_b} 	 

\newcommand{\md}{Md}
\newcommand{\td}{Td}
\newcommand{\ad}{Ad}

\def\barray{\begin{array}} 
\def\earray{\end{array}}
\def\be{\begin{equation}}
\def\ee{\end{equation}}
\def\ben{\begin{equation} \nonumber}
\def\een{\end{equation}}
\def\ban{\begin{eqnarray*}}
\def\ean{\end{eqnarray*}}
\def\ba{\begin{eqnarray}}
\def\ea{\end{eqnarray}}

\def\neff{$N_{\rm eff}$}
\def\msum{$\sum m_\nu$}
\def\ms{$m_s$}

\def\({\left(}
\def\){\right)}

\begin{document}

\title{$\nu\Lambda$CDM: Neutrinos help reconcile Planck with the Local Universe}
\author{Mark Wyman}
\email{markwy@oddjob.uchicago.edu}
\author{Douglas H. Rudd}
\author{R. Ali Vanderveld}
\author{Wayne Hu }
\affiliation{Kavli Institute for Cosmological Physics,  Department of Astronomy \& Astrophysics,  Enrico Fermi Institute, University of Chicago, Chicago, Illinois 60637, U.S.A}

\begin{abstract}
Current measurements of the low and high redshift  Universe are in tension if we restrict ourselves to the standard six parameter
model of flat $\Lambda$CDM. This tension has two parts. First, the Planck satellite data suggest a higher normalization of matter perturbations than local measurements of galaxy clusters. Second, the expansion rate of the Universe today, $H_0$, derived from local distance-redshift measurements is significantly higher than that inferred
 using the acoustic scale in galaxy surveys and the Planck data as a standard
ruler. The addition of 
a sterile neutrino species changes the acoustic scale and brings the two into agreement;
meanwhile,  adding mass to the active neutrinos or to a sterile neutrino can suppress the growth of structure, bringing the cluster data into better concordance as well. For our fiducial dataset combination, with statistical errors for clusters, a model with a massive sterile neutrino
shows 3.5$\sigma$  evidence for a non-zero mass and an even stronger rejection of the minimal model. 
A model with massive active neutrinos and a massless sterile neutrino is similarly preferred.
An eV-scale sterile neutrino mass -- of interest for short baseline and reactor anomalies -- is well within the allowed range.  We caution that 1) unknown astrophysical systematic errors in any of the
data sets could weaken this conclusion, but they would need to be several times the known errors to eliminate the tensions entirely; 2) the results we find are at some variance with analyses that do not include cluster measurements; and 3) some tension remains among the datasets even when new neutrino physics is included.
\end{abstract}

\maketitle

Neutrinos are one of the most elusive constituents of the standard model of particle physics.
They interact only via the weak force and are nearly massless. In the standard picture, there are three neutrino species
with a summed mass 
that solar and atmospheric oscillation observations bound to be above 0.06 eV (e.g. 
\cite{GonzalezGarcia:2012sz}).
 However, anomalies in short baseline and reactor neutrino experiments suggest that there may be
one or more additional eV scale massive sterile neutrinos (see refs. \cite{Conrad:2013mka,Abazajian:2012ys} for reviews). 

Meanwhile,  cosmological observations
have established a standard model of cosmology -- often called inflationary 
$\Lambda$CDM.    With only six basic parameters, its most minimal incarnation
can explain a wide range of phenomena, from light element abundances, through the cosmic microwave background (CMB) anisotropy and large scale structure, the formation and statistical properties
of dark matter halos that host galaxy clusters to the current expansion history and cosmic acceleration.  Precise new
data allow us to test if the subtle effects of eV scale neutrinos and partially populated
sterile species are also present.

Interestingly, the Planck satellite \cite{Ade:2013zuv} has
recently exposed potential tension 
between the early and late time
observables in the minimal six parameter  model. 
In particular, Planck finds a larger and more precisely measured
matter density at recombination than previous data.
   This relatively small change
at high redshift
cascades into more dramatic implications for observables
today (e.g.~\cite{Hu:2004kn}): the current expansion rate, $H_0$, decreases and 
the amount of cosmological structure increases.   These changes are in  2-3$\sigma$ tension with
direct observations of $H_0$  {\cite{Riess:2011yx} and the abundance
of galaxy clusters  \cite{Vikhlinin:2008ym}, respectively.   Meanwhile, agreement with distance
measures from baryon acoustic oscillations (BAO)  \cite{Anderson:2012sa, Padmanabhan:2012hf, Blake:2011en} suggest that the former cannot be 
resolved by having evolving dark energy modify the recent expansion history.  

Neutrinos offer a possible means of bringing these observations into concordance.  Sterile neutrinos change the expansion rate at recombination
and hence the calibration of the standard ruler with which CMB and BAO observations
infer distances (e.g.~\cite{Ade:2013zuv}). 
When either the sterile or active species are
massive, their free streaming reduces the 
amount of small scale clustering today and hence the tension with cluster measurements. 
In the simplest case, we can think of this modification as adding a single, massive sterile neutrino to the standard model. 

\smallskip{\it Models and Data.--}
The minimal 6 parameter flat $\Lambda$CDM model is defined by
$\{\Oc h^2, \, \Ob h^2, \, \tau, \theta_A, A_{\rm S}, n_s \}$, where $\Oc h^2$ defines the cold dark matter (CDM) density, $\Ob h^2$  the baryon density, $\tau$ the Thomson optical depth to reionization, $\theta_A$ the angular acoustic scale at recombination, $A_s$ the amplitude of the initial curvature power spectrum at $k = 0.05$ Mpc$^{-1}$, and $n_s$ its spectral index.  With precise constraints on these
parameters from  CMB  data at high redshift, all other low redshift observables are
precisely predicted: importantly 
the Hubble constant, $H_0 = 100 h$ km/s/Mpc, the present total matter density $\Om$, and the  rms amplitude of linear fluctuations today on the
$8 h^{-1}$Mpc scale, $\sigma_8$. 

Conflict between
these predictions and actual measurements may suggest a non-minimal model. In this context, we
consider 3 new neutrino parameters: $N_{\rm eff}$, $\sum m_\nu$, and $m_s$. 
We define $N_{\rm eff}$, the effective number of relativistic species, via the relativistic energy density at high redshift
\be
\rho_{\rm r} = \rho_\gamma + \rho_\nu = \left[ 1 + \frac{7}{8} \(\frac{4}{11}\)^{4/3}  \hspace{-6pt} N_{\rm eff} \right] \rho_\gamma.
\ee
In the minimal model $N_{\rm eff}=3.046$. Any value of $N_{\rm eff}$ larger than
this fiducial value 
corresponds to a greater expansion rate in the early Universe, consistent with the presence of some extra density of relativistic particles, which includes  neutrinos beyond the 3 known ``active"  species. 
Next, $\sum m_\nu$ denotes the summed mass of the active neutrinos. It is at least $0.06$ eV, from  mass squared splittings in solar and atmospheric oscillations,
but in principle could be larger if the species are nearly degenerate in mass. 
We call the model with $N_{\rm eff}=3.046$, $\sum m_\nu=0.06$eV the ``minimal neutrino"
(\mnu) mass model.

 Finally, we introduce
an effective mass $m_s$ for the 4th, mostly sterile, species by requiring that the total neutrino contribution
to the energy density today is given by 
\be
\label{omegan}
({94.1 \, \rm eV}) \Omega_\nu h^2 = {(3.046/3)^{3/4} \sum m_\nu+m_s}.
\ee
We do not study all three extra parameters simultaneously, but instead vary \neff\ together with either \msum\
or \ms\ -- see Table \ref{tab:models}.
When we allow $m_s$ to vary we set $\sum m_\nu = 0.06$eV
and call it the ``sterile neutrino" (\snu) mass model. Similarly, we explore an 
``active neutrino" (\anu) model, allowing $\sum m_\nu$ to vary with the masses assumed to be degenerate and setting $m_s=0$.
  We define the total non-relativistic matter density today as $\Om = \Oc+\Ob+\Omega_\nu$.

\begin{table}[t!]
\caption{\footnotesize Models and data combinations studied.}
\begin{center}
\begin{tabular}{l || c | c | c | c}
 Model  & $\Lambda$CDM (6)  & $N_{\rm eff}$ & $\sum m_\nu$ & $m_{s}$ \\ \hline
M(inimal)$\nu$& \checkmark & 3.046 & 0.06eV  & 0  \\
\hline
S(terile)$\nu$&\checkmark & \checkmark & 0.06eV & \checkmark\\
\hline
A(ctive)$\nu$& \checkmark & \checkmark & \checkmark & 0  \\
\end{tabular}
\label{tab:models}
\vspace{0.2in}
\begin{tabular}{l || c | c | c  }
Data & M(inimal)d &T(ension)d & A(ll)d  \\ \hline
Planck \cite{Ade:2013zuv} +WMAP P. \cite{Bennett:2012fp} & \checkmark & \checkmark & \checkmark \\
$H_0$  \cite{Riess:2011yx} & &\checkmark & \checkmark \\
BAO \cite{Anderson:2012sa, Padmanabhan:2012hf, Blake:2011en} & &\checkmark &\checkmark  \\
X-ray Clusters \cite{Vikhlinin:2008ym} &&\checkmark&\checkmark\\
SNe (Union2) \cite{Suzuki:2011hu} &&&\checkmark \\
High-$\ell$ CMB \cite{Reichardt:2011yv,Keisler:2011aw,Das:2013zf} &&& \checkmark \\
\end{tabular}
\label{tab:data}
\end{center}
\end{table}

Note that $m_s$ is not the true mass of a new neutrino-like particle, but rather encapsulates
both the particle's mass and how this species was populated in the early universe.
This effective mass is typically related to the true mass in one of two ways. If the extra sterile neutrino species
are thermally distributed, we have $m_s^{\rm T} = (\Delta N_{\rm eff})^{-3/4} m_s,$
where we have defined $\Delta N_{\rm eff} = N_{\rm eff} - 3.046 \equiv (T_\nu/T_s)^3$. Alternatively,
if the new sterile neutrino(s) are distributed proportionally to the active neutrinos due to oscillations, we have, following Dodelson and Widrow \cite{Dodelson:1993je},
$m_s^{\rm DW} =( \Delta N_{\rm eff})^{-1} m_s$.
Since the effective parameter that enters the cosmological analysis is the same in both cases, the choice only impacts the interpretation and external priors.  For the latter, we take a $m_s^{\rm DW} < 7{\rm eV}$ prior to prevent trading 
 very massive neutrinos with CDM -- a degeneracy which is not of interest for eV scale
neutrino physics. Note that we use this condition to set an allowed prior range in the $m_s$-$\Delta N_{\rm eff}$ plane.
We will otherwise take flat priors on the separate $m_s$ and $\Delta N_{\rm eff}$ parameters.


To explore constraints on these parameters given the various cosmological data sets,
we sample their posterior probability with the Monte Carlo Markov Chain technique using
the  CosmoMC code \cite{Lewis:2002ah} for the various data sets summarized in Table \ref{tab:data}. Common to all sets is the CMB temperature data from the Planck satellite \cite{Ade:2013zuv} together with polarization data from the WMAP satellite \cite{Bennett:2012fp}, dubbed the ``minimal" dataset (\md).   Here we marginalize the standard foreground nuisance parameters provided
by Planck.   Note that CosmoMC in practice uses an approximation to the acoustic
scale $\theta_{\rm MC}\approx \theta_A$ and uses $\ln A= \ln(10^{10}A_S)$.

Next, we add datasets that reveal the presence of tension with the \mnu\ model. These are the  $H_0$ inference from the maser-cepheid-supernovae  distance ladder  \cite{Riess:2011yx},  
BAO measurements 
\cite{Anderson:2012sa, Padmanabhan:2012hf, Blake:2011en} and the X-ray cluster abundance \cite{Burenin:2012uy} \footnote{The cluster
abundance is included in the likelihood analysis via the likelihood code released simultaneously with \cite{Burenin:2012uy}}.   
We call this combined dataset the ``tension" dataset (\td).  This is the minimal set of data required to expose tension.  The BAO data, which also measure the low redshift distance-redshift relation,  prevent explaining $H_0$ with smooth changes
in the expansion history toward phantom equations of state
\cite{Ade:2013zuv}. Thus, we include the BAO data in the ``tension" dataset because 
it confirms the existence of tension, not because it is itself in tension with Planck. Conversely, clusters alone might be explained by exotic dark energy that reduces the linear
growth rate; but when combined with these distance measurements, the data point to neutrinos instead.

 For the cluster data, we also separately test a systematic 9\% increase in the mass calibration of local
clusters \cite{Vikhlinin:2008ym} to show the shift in some of our statistics. Use of such a shift was proposed by the authors of \cite{Vikhlinin:2008ym}; its size was based on a variety of X-ray, optical, Sunyaev-Zel'dovich, and lensing mass observables (see e.g.~\cite{Rozo:2013hha}).
Finally, we add the Union2 compilation of type Ia supernovae \cite{Suzuki:2011hu}
and high resolution CMB data \cite{Dunkley:2013vu}
from the ACT \cite{Das:2013zf} and SPT \cite{Reichardt:2011yv,Keisler:2011aw} telescopes in the ``all" dataset (\ad).

\smallskip{\it Results.--}
We start with the basic minimal neutrino model and minimal Planck-WMAP dataset case (\mnu-\md) shown in 
Tab.~\ref{tab:neutrino} (column 1).   From the fundamental chain parameters, we can
derive the posterior probability distributions for  two auxiliary parameters, $H_0$ and
$\som=\sigma_8 (\Omega_m/0.25)^{0.47}$ -- see Fig.~\ref{fig:tension}.   The latter effectively controls the local cluster abundance.       
Very little overlap exists between the \mnu-\md\ predictions for these local
observables and the measurements (68\% confidence bands).  Even adding a 9\% systematic shift in the
cluster masses  is insufficient to bring about concordance.  
 
These predictions
depend on our assumptions about neutrinos.
 The presence of extra relativistic species in the early Universe alters the expansion rate
 and thus the physical length scale associated with both the CMB and the BAO. Allowing \neff\ to vary changes this scale and broadens the allowed range for $H_0$.   In Fig.~\ref{fig:tension} (bottom), we see that in the  \snu\ case, the $H_0$ posterior implied by  \md\  broadens to include substantial overlap
with the measurements.    A similar broadening occurs for the \anu\ case.

\begin{figure}[t!]
\includegraphics[width=\columnwidth]{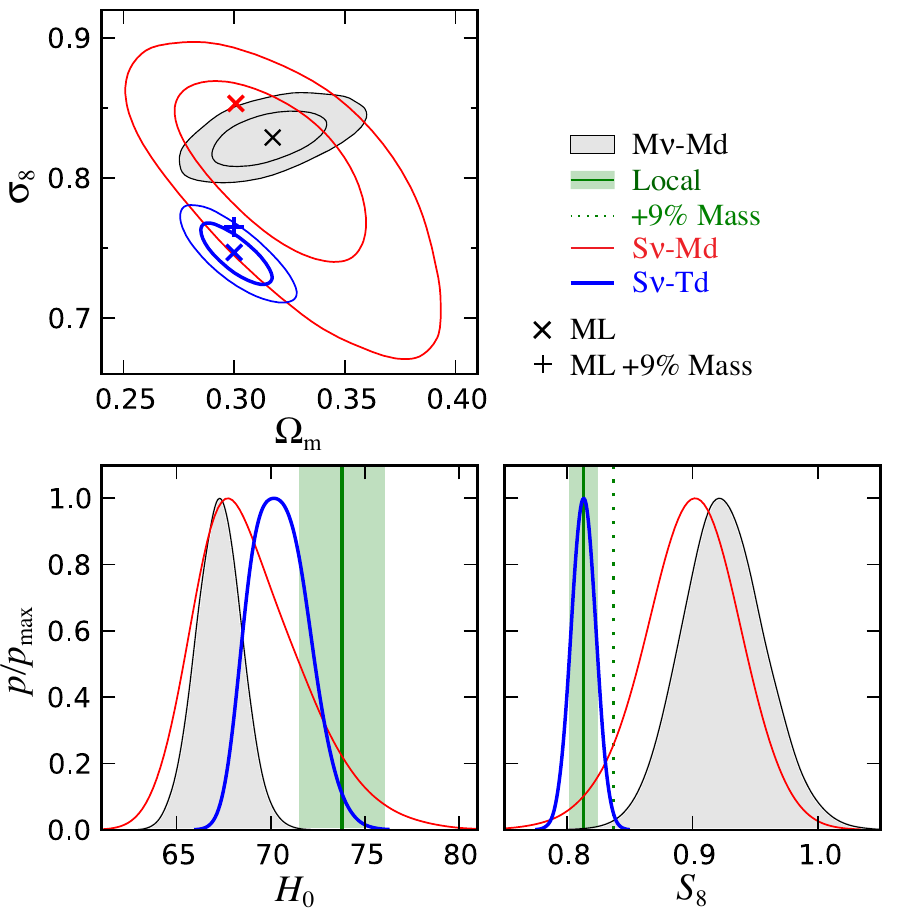}
\caption{\footnotesize Tensions between datasets and their neutrino alleviations.  Black, red, and blue
curves represent the \mnu-\md, \snu-\md, and \snu-\td\ model-data combinations respectively.
{\it Bottom:} $H_0$ and $\som$ posteriors (curves) vs.\ local measurements (bands, 68\% CL).  Lack of overlap
in \mnu-\md\ is alleviated in \snu-\md\ leading to better concordance in \snu-\td.  The dashed line
shows the change in $\som$ from the 9\% cluster mass offset.  {\it Top:} $\sigma_8$ and $\Om$ 68\% and 95\% confidence regions.  Neutrino parameters open a direction mainly orthogonal to $S_8$.  ``$\times$" marks ML models; ``+" shows its shift
 for a 9\% cluster mass offset. \anu\ model results are similar.}
\label{fig:tension}
\end{figure}

\begin{figure}
\includegraphics[width=\columnwidth]{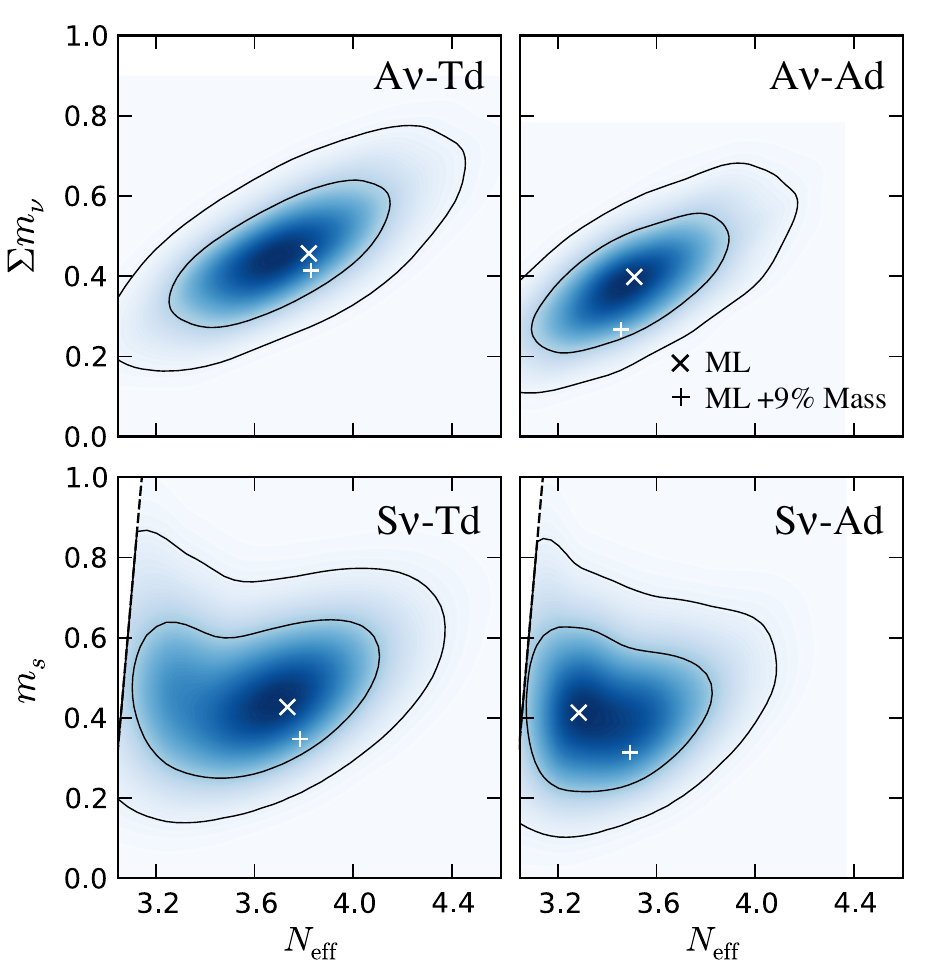}
\caption{\footnotesize Neutrino mass and effective number constraints, labelled as in Fig.~\ref{fig:tension}
($\times$ indicates the ML model, $+$ its shift from a 9\% cluster mass increase).
 {\it Bottom:} \snu\ sterile case for \td\ (left) and \ad\ (right).  The region excluded by the $m_s^{\rm DW} < 7{\rm eV}$ prior is left of the dashed line. 
{\it Top:} \anu\ active case for \td\ (left) and \ad\ (right).   In all cases the minimal  $\sum m_\nu=0.06$eV,
\neff$=3.046$ and $m_s=0$ is highly excluded.}
\label{fig:neutrinopar}
\end{figure}

\begin{table*}
\caption{\footnotesize Summary of posterior statistics.  
 $\Omega_m$, $H_0$ and $\som$ are derived parameters and $2\Delta\ln{\cal L}$ gives the  likelihood of the ML model of the non-minimal neutrino model relative to the minimal \mnu\ model with the same dataset. Upper limits are 68\% CL.}
\begin{center}
\begin{tabular}{|l||c|c|c|c|c|c|c|}
\hline
Data  & \multicolumn{3}{|c|}{\md}&\multicolumn{2}{|c|}{\td}&\multicolumn{2}{|c|}{\ad}\\ \hline
Model & \mnu& \snu& \anu& \snu& \anu& \snu& \anu\\
$2\Delta\ln {\cal L}$& --& 0.5& 0.9& 15.5& 14.1& 11.9& 9.7\\ \hline
$100\Omega_b h^2$ & $2.204 \pm 0.028$ & $2.236 \pm 0.036$ & $2.222 \pm 0.046$ & $2.272 \pm 0.027$ & $2.275 \pm 0.028$ & $2.272 \pm 0.027$ & $2.273 \pm 0.028$  \\
$\Omega_c h^2$ & $0.1199 \pm 0.0027$ & $0.1263 \pm 0.0052$ & $0.1255 \pm 0.0053$ & $0.1210 \pm 0.0050$ & $0.1229 \pm 0.0044$ & $0.1183 \pm 0.0040$ & $0.1196 \pm 0.0038$  \\
$100\theta_{\rm MC}$ & $1.0413 \pm 0.0006$ & $1.0406 \pm 0.0007$ & $1.0407 \pm 0.0008$ & $1.0412 \pm 0.0007$ & $1.0409 \pm 0.0007$ & $1.0414 \pm 0.0006$ & $1.0413 \pm 0.0007$  \\
$\tau$ & $0.090 \pm 0.013$ & $0.095 \pm 0.015$ & $0.094 \pm 0.014$ & $0.096 \pm 0.015$ & $0.096 \pm 0.015$ & $0.096 \pm 0.014$ & $0.096 \pm 0.015$  \\
$n_s$ & $0.9604 \pm 0.0072$ & $0.9748 \pm 0.0148$ & $0.9721 \pm 0.0175$ & $0.9857 \pm 0.0120$ & $0.9939 \pm 0.0097$ & $0.9798 \pm 0.0108$ & $0.9877 \pm 0.0096$  \\
$\ln A$ & $3.089 \pm 0.025$ & $3.116 \pm 0.031$ & $3.110 \pm 0.033$ & $3.107 \pm 0.031$ & $3.109 \pm 0.031$ & $3.101 \pm 0.030$ & $3.100 \pm 0.032$  \\
$N_{\rm eff}$ & --  & $3.56 \pm 0.31$ & $3.44 \pm 0.38$ & $3.61 \pm 0.31$ & $3.72 \pm 0.29$ & $3.44 \pm 0.23$ & $3.51 \pm 0.26$  \\
$\Sigma m_\nu$,$m_s$ & --  & $<0.34$ & $<0.32$ & $0.48 \pm 0.14$ & $0.46 \pm 0.12$ & $0.44 \pm 0.14$ & $0.39 \pm 0.11$  \\ \hline
$\Omega_m$ & $0.316 \pm 0.017$ & $0.322 \pm 0.028$ & $0.331 \pm 0.050$ & $0.301 \pm 0.010$ & $0.299 \pm 0.011$ & $0.298 \pm 0.010$ & $0.296 \pm 0.010$  \\
$H_0$ & $67.3 \pm 1.2$ & $69.0 \pm 2.8$ & $67.9 \pm 4.5$ & $70.5 \pm 1.5$ & $70.9 \pm 1.4$ & $70.0 \pm 1.2$ & $70.4 \pm 1.4$  \\
$\som$ & $0.925 \pm 0.033$ & $0.899 \pm 0.038$ & $0.908 \pm 0.036$ & $0.813 \pm 0.010$ & $0.815 \pm 0.009$ & $0.813 \pm 0.010$ & $0.815 \pm 0.009$  \\
\hline
\end{tabular}
\end{center}
\label{tab:neutrino}
\end{table*}

Allowing part of the matter 
 to be composed of neutrinos with eV scale masses suppresses the growth of structure
 below their free-streaming length.   This allows $\sigma_8$ to 
 be substantially lower and still be compatible with the \md\ CMB datasets (see Fig.~\ref{fig:tension}).   However, since the CDM component $\Omega_c h^2$ is
 well constrained independently, adding neutrinos increases $\Omega_m$, leading
 to a  less pronounced modification to the cluster observable (see Fig.~\ref{fig:tension},
bottom right and top panels).   Also, raising \neff\ to reduce the $H_0$ tension  requires
 an increase in the tilt $n_s$ to compensate for the reduction of power in the CMB damping tail, which further reduces the impact (see e.g.~\cite{Vanderveld:2012yp} Fig.  3).
  Nonetheless the overlap between the posterior of the \md\ dataset and the measurements is now visible for the \snu\ model, whereas it was negligible with the
 \mnu\  model.    Furthermore, a 9\% shift in cluster masses now brings the observations
 into reasonable concordance.    Slightly more tension remains in the \anu\ case because 
spreading the mass among three species gives lower true masses for each. 
Including the BAO and $H_0$ data also somewhat enhance the residual tension
with high mass \cite{Verde:2013cqa}.
 
 A joint analysis of the \td\ data set supports these conclusions (see Tab.~\ref{tab:neutrino}).  
 For the \snu\  model, the minimal neutrino values of $m_s=0$ and \neff$=3.046$ are 
 individually disfavored at 3.5$\sigma$ and 2$\sigma$ respectively.  Fig.~\ref{fig:neutrinopar}
 shows that the joint exclusion is even stronger, with the minimal \neff\ at $m_s=0$  
 rejected at high confidence.
 The maximum likelihood (ML) \snu\ model has a $2\Delta\ln {\cal L}= 15.5$ with two extra parameters  ($m_s=0.43$eV and \neff$=3.73$) over that of the \mnu\ model.
 Note that these two parameters combine to imply an actual ML
 mass $m^{\rm DW}_s=0.62$eV.
   For the  \anu-\td\ case, 
 the minimal $\sum m_\nu$ and \neff\ are disfavored  at 3.4$\sigma$ and $2.3\sigma$
 respectively with 
  $2\Delta \ln {\cal L}= 14.05$ ($\sum m_\nu=0.46$eV, \neff$=3.82$).
   
 Including all of the data with \ad\ reduces these preferences somewhat (see Tab.~\ref{tab:neutrino} and Fig.~\ref{fig:neutrinopar}).   This is mainly due to the high resolution CMB data which can break degeneracies between parameters like 
 \neff\ and $n_s$.   But the preference for non-minimal masses remains: $3.2\sigma$
 and
 $3\sigma$ evidence (with improvements of
  $2\Delta \ln {\cal L}= 11.9$ and $9.7$) for the \snu\ and \anu\ cases respectively.
 
 With a 9\% cluster mass offset, lower neutrino masses are preferred.   
 For example, in the \snu-\td\ case the ML model shifts from  $m_s=0.43$eV to
 $0.35$eV with ML improvement of \snu\ over \mnu\ of 
 $2\Delta \ln {\cal L}=9.6$.   For the \anu-\td\ case it shifts from $\sum m_\nu=0.46$eV to
 $0.41$eV with $2\Delta \ln {\cal L}=8.4$.
 Other cases are shown in Fig.~\ref{fig:neutrinopar} and all are within the 68\% joint CL regions.

\smallskip{\it Discussion.--}  Taken at face-value, these results indicate $\sim 3\sigma$ statistical evidence
for non-minimal neutrino parameters, especially in their masses, which simultaneously brings concordance
in the CMB, BAO,
$H_0$, and  cluster data. The addition of other datasets, such as supernovae or high-$\ell$ CMB measurements,
refine but do not qualitatively change this conclusion.   

Conversely, unknown systematic errors in any of the
\td\ data sets could alter our conclusions substantially.  For Planck these include the modeling of foregrounds and instrumental effects, especially at high multipole, and for $H_0$ the calibration of the supernova distance ladder.  The preference for high neutrino mass(es)
is mainly driven by the cluster data set (cf.~Ref.\ \cite{Verde:2013cqa} who find upper limits without clusters).  As such, the best fitting parameter values we find are in mild tension with results from combinations of datasets that exclude clusters. However,
the improvement in the agreement with the cluster data is sufficiently strong to more than compensate, in a likelihood maximization sense, for this slight worsening of the fit to the other data. In light of these various concerns -- especially the remaining tension even with new neutrino physics included -- a more thoroughgoing model-selection analysis of these data
will certainly be warranted in the future, especially as the systematic errors in each of the datasets become better quantified.
Regardless, if future data or analyses lead to increased mass estimates for the clusters,
that change would weaken the preference we find. However,
 the preference can only be eliminated if the systematic shift is roughly triple
the 9\% estimate.    As mention before, this cluster mass calibration error estimate comes from comparing
a variety of X-ray, optical, Sunyaev-Zel'dovich, and lensing observables (see e.g.~\cite{Rozo:2013hha} for a recent assessment).  We also note that the Planck Sunyaev-Zeldovich  cluster results are consistent with the dataset we have used, and their analysis of neutrino physics agrees with ours where the two overlap; however, the Planck collaboration did not directly test the neutrino models that we have used in their analyses \cite{Ade:2013lmv}.

Other cosmological data sets can also cross check these conclusions.   Indeed, there
is mild tension with the shape of  galaxy power spectra
\cite{Riemer-Sorensen:2013jsa,Giusarma:2013wsa}
but these come with their own astrophysical systematics in the interpretation of galaxy bias, and those systematics are more difficult to address than
 those affecting cluster mass estimates.   
In the future, weak lensing of the CMB and galaxies 
should definitively test this result.

{
\smallskip{\em Acknowledgments.--}  We thank E.~Rozo, A.~Zablocki, and S.~Dodelson for helpful discussions. 
This work was supported at the KICP through grants NSF PHY-0114422 and NSF PHY-0551142  and an endowment from the Kavli Foundation. MW and WH were additionally supported by U.S.~Dept.\ of Energy contract DE-FG02-90ER-40560. 
This work made use of computing resources provided by the Research Computing Center at the University of Chicago.}\par}

\bibliography{bibneut}

\end{document}